\documentclass[11pt]{article}
\usepackage{amsmath,amsfonts}
\usepackage{tikz}
\usetikzlibrary{trees,er,snakes,shapes,mindmap}
\allowdisplaybreaks
\def \beq  {\begin{equation}}
\def \eeq  {\end{equation}}
\def \beqar {\begin{eqnarray}}
\def \eeqar {\end{eqnarray}}
\def\sqr#1#2{{\vcenter{\vbox{\hrule height.#2pt
\hbox{\vrule width.#2pt height#1pt \kern#1pt
\vrule width.#2pt}\hrule height.#2pt}}}}

\def\la {{\langle}}
\def\ra {{\rangle}}

\def\vf {{\varphi}}

\def\Tr {{\rm Tr}}

\def\ba {\bar{a}}
\def\bD {\bar{D}}
\def\bA {\bar{A}}

\def\bx {\bar{x}}
\def\by {\bar{y}}

\def\vf {{\varphi}}

\def\del {\partial}
\def\bdel{\bar{\partial}}

\def\bz {{\bar{z}}}

\def\C {{\cal C}}

\def\M{{\cal M}}

\def\half{\textstyle{1\over 2}}

\begin{document}
\fontfamily{cmr}\fontsize{11pt}{15pt}\selectfont
\def \CMP {{Commun. Math. Phys.}}
\def \PRL {{Phys. Rev. Lett.}}
\def \PL {{Phys. Lett.}}
\def \NPBProc {{Nucl. Phys. B (Proc. Suppl.)}}
\def \NP {{Nucl. Phys.}}
\def \RMP {{Rev. Mod. Phys.}}
\def \JGP {{J. Geom. Phys.}}
\def \CQG {{Class. Quant. Grav.}}
\def \MPL {{Mod. Phys. Lett.}}
\def \IJMP {{ Int. J. Mod. Phys.}}
\def \JHEP {{JHEP}}
\def \PR {{Phys. Rev.}}
\def \JMP {{J. Math. Phys.}}
\def \GRG{{Gen. Rel. Grav.}}
\begin{titlepage}
\null\vspace{-62pt} \pagestyle{empty}
\begin{center}
\rightline{CCNY-HEP-13/4}
\rightline{July 2013}
\vspace{1truein} {\Large\bfseries
On the Gauge-invariant Functional Measure for
}\\
\vspace{6pt}
\vskip .1in
{\Large \bfseries  Gauge Fields on ${\mathbb {CP}}^2$}\\
\vskip .1in
{\Large\bfseries ~}\\
{\large\sc V.P. Nair}\\
\vskip .2in
{\itshape Physics Department\\
City College of the CUNY\\
New York, NY 10031}\\
~\\
and\\
~\\
{\itshape 
Niels Bohr International Academy\\
Niels Bohr Institute\\
Blegdamsvej 17\\
DK-2100 Copenhagen\\
Denmark}
\vskip .1in
\begin{tabular}{r l}
E-mail:
&{\fontfamily{cmtt}\fontsize{11pt}{15pt}\selectfont vpn@sci.ccny.cuny.edu}
\end{tabular}

\fontfamily{cmr}\fontsize{11pt}{15pt}\selectfont
\vspace{.8in}
\centerline{\large\bf Abstract}
\end{center}
We introduce a general parametrization for
nonabelian gauge fields on the four-dimensional space
${\mathbb{CP}}^2$. 
The volume element for the gauge-orbit space or the space of physical configurations
is then investigated. The leading divergence in this volume element is obtained in terms
of a higher dimensional Wess-Zumino-Witten action, which has previously been studied in the context of K\"ahler-Chern-Simons theories.
This term, it is argued,
implies that one needs to introduce a dimensional parameter to specify the integration measure,
a step which is a nonperturbative version of the well-known dimensional transmutation
in four-dimensional gauge theories.

\end{titlepage}

\pagestyle{plain} \setcounter{page}{2}
\section{Introduction}

The importance of the gauge-orbit space needs no emphasis given that Yang-Mills theories are the foundational paradigm for the interactions of fundamental particles.
The relevant space over which the functional integration for such theories is carried out
is the space of gauge potentials ($ {\cal A}$) modulo the space of all gauge transformations 
which are fixed to be identity at one point on the spacetime manifold (${\cal G}_*$). In particular, 
the measure of integration is the volume element of this gauge-orbit space 
$\C = {\cal A }/ {\cal G}_*$, which is, equivalently, the space of
physical field configurations \cite{singer}.

This volume element can be calculated exactly for gauge fields in two dimensions in terms of a Wess-Zumino-Witten (WZW) action \cite{gawe}. It plays a role in the Chern-Simons-WZW relationship \cite{witten} and, albeit indirectly,  in the solution of Yang-Mills theory on Riemann surfaces \cite{2dYM}.
It can be incorporated into a Hamiltonian formalism for (2+1)-dimensional Yang-Mills theories leading to string tension calculations and insight into the mass gap \cite{{KKN}, {nair-trento1}}, including supersymmetric cases \cite{AN1}.

The situation for four-dimensional gauge theories has been much less clear.
Gauge-fixing and the Faddeev-Popov procedure construct this volume element
for a local section of ${\cal A}$ viewed as a ${\cal G}_*$-bundle over 
$\C$; this is adequate for the perturbative calculations, but does not really give
any insight into anything beyond that. The volume element for the gauge-orbit space
for four dimensional gauge fields is the subject of this paper.
The calculations in lower dimensions
utilized the possibility that one could view two-dimensional space as a complex manifold, which then led to a parametrization of fields which was very suitable for the calculation of the volume element for $\C$. There is no natural complex structure for ${\mathbb{R}}^4$ since any choice of complex coordinates would not be 4d-rotationally invariant (or Lorentz invariant  with Minkowski signature). One could consider a twistor space version which would include the set  of all local complex structures. However, a simpler situation is obtained with ${\mathbb{CP}}^2$, which is a complex K\"ahler manifold. The standard metric for this space is the Fubini-Study metric given,
in local coordinates $z^a$, $\bz^{\bar a}$, $a= 1,2$, $\ba = {\bar 1}, {\bar 2}$, by
\beq
ds^2 = {dz^a \, d\bz^{\bar a} \over (1+ z\cdot \bz/R^2)}
- {\bz \cdot dz \, z\cdot d\bz \over R^2 (1+ z\cdot \bz /R^2)}
\label{0a}
\eeq
where we have also included a scale parameter for the coordinates. As the parameter
$R \rightarrow \infty$, the metric becomes that of flat space (although there are some global issues which will not be important for us).
This is, therefore, an interesting space to consider, being endowed with a complex structure and with a suitable limit to the flat four-dimensional space. So, in this paper, we will consider gauge fields on
${\mathbb {CP}}^2$.

In the next section we will introduce a suitable parametrization for
gauge fields on ${\mathbb{CP}}^2$ and identify the gauge-invariant variables of the problem.
We will then proceed to the evaluation of the leading divergent term in the functional integration measure.
This is shown to be given by a higher dimensional generalization of the WZW action.
The functional integration for gauge fields in four dimensions, it is well known, 
 should show dimensional transmutation with a freely specifiable dimensional parameter characterizing the theory.
 In the last section, we
argue that the leading divergence in the calculation of the volume element for
$\C$ introduces just such a parameter, which is, effectively, 
a nonperturbatively defined version of the
$\Lambda$-parameter of QCD.
The main result of the paper is then contained in equations (\ref{51}) and (\ref{52}),
which give the definition of the functional integral with the measure defined in terms the gauge-invariant variables.
The computation of the subleading and finite terms in the Jacobian of the transformation to the gauge-invariant variables and the extensions of the result to supersymmetric theories 
are briefly alluded to in the discussion section; they
are
interesting directions to explore in future.

\section{The volume element for the gauge orbit space}

\subsection{Parametrization of fields}

We will begin with a suitable parametrization of the gauge fields on ${\mathbb {CP}}^2$.
This space may be thought of as the group coset $SU(3)/ U(2)$.
Thus functions, vectors, etc. on this space may be realized in terms of the Wigner functions
$\la R, A \vert {\hat g} \vert R, B\ra$ which are the representation of an $SU(3)$ element
$g$ in a general irreducible representation labeled as $R$. For the defining fundamental representation, we take $g$ to be a $3\times 3$ unitary matrix of unit determinant and
it can be parametrized as $g = \exp ({i t_a \, \vf^a})$, where $t_a$ form a basis for hermitian 
$3\times 3$ matrices, with $\Tr\,( t_a t_b) = \half \, \delta_{ab}$, and $\vf^a$ are the coordinates
 for $SU(3)$.
In terms of the functions $\la R, A \vert {\hat g} \vert R, B\ra$, the states on the right, namely, $\vert R, B\ra$ must be so chosen as to give the correct transformation property under $U(2) \in SU(3)$.
Notice that, for ${\mathbb{CP}}^2$, $SU(3)$ plays the role of the Poincar\'e group and $U(2)$ plays the role of the Lorentz group; so vectors, tensors, etc., must be characterized by their transformation property
under $U(2)$. We will refer to the $SU(2)$ part of $U(2)$ as isospin (denoted by $I$) and
the $U(1)$ part of $U(2)$ as hypercharge (denoted by $Y$).
Specifically, we take the $SU(2)$ to be generated by $t_1, \, t_2,\, t_3$ and the hypercharge
to correspond to $2\,t_8/\sqrt{3}$.
For functions on $\mathbb{CP}^2$, which must be invariant under $U(2)$, we need states with
$Y =0$ and $I =0$. For vectors, we need 
an $SU(2)$ doublet ($I = \half$ representation). A general $SU(3)$ representation is of the form
$T^{a_1 a_2 \cdots a_p}_{b_1 b_2 \cdots b_q}$, $a_i,\, b_j = 1,2, 3$, which may be labeled as
$(p,q)$. These are totally symmetric in all $a_i$'s and totally symmetric in all
$b_j$'s with the trace (or contraction between any choice of upper and lower indices) vanishing.
The value of hypercharge is given by
\beq
Y = \left\{ \begin{array}{r c l}
1/3 &~& a_i = 1,2\\
-2/3&~& a_i = 3\\
- 1/3 &~& b_i = 1, 2\\
2/3&~& b_i =3\\
\end{array}\right.
\label{1}
\eeq
For the derivative operators on
${\mathbb {CP}}^2$ we can use a subset of the right translation operators
$R_a$ defined by
\beq
R_a \, g = g \, t_a
\label{2}
\eeq
More explicitly, we can write
\beq
g^{-1} \, dg = - i t_a \, E^a_i \, d\vf^i,
\hskip .2in R_a = i (E^{-1})^i_a \, {\del \over \del \vf^i}
\label{3}
\eeq
For $\mathbb{CP}^2$, the derivatives will be taken as
$\nabla_1 = R_4 + i R_5$ and $\nabla_2 = R_6+i R_7$, and $\nabla_{\bar 1} =
R_4 -i R_5, \, \nabla_{\bar 2} = R_6 -i R_7$. The $\nabla$'s correspond to derivatives in
the tangent frame, $\nabla_i = i (e_i^{-1})^\mu (\del /\del x^\mu)$ in terms of the usual
local coordinates, $e$'s being the frame fields.
(The group theoretic approach we use for ${\mathbb{CP}}^k$ spaces, with derivatives given by
$R_a$, is essentially along the lines of
\cite{KNqhe}; it is also similar to what was done for gauge fields on 
${\mathbb{CP}}^1$ in \cite{AN2}.)

The operators $\nabla_i$ form an $SU(2)$ doublet with $Y =1$,
$\nabla_{\bar i}$ are again an $SU(2)$ doublet with $Y = -1$. A gauge field $A$
is to be added to these operators, so we need an $SU(2)$ doublet with
$Y= 1$ for $A_i$, and an $SU(2)$ doublet with $Y = -1$ for
$A_{\bar i}$. This corresponds to the states of the form
$T^{i 3 3 \cdots 3}_{3 3 \cdots 3}$ which give $Y = 1$ for $p =q$, 
and $Y =- 1$ for $p = q+3$. Likewise, $T^{3 3 \cdots 3}_{i 3 3 \cdots 3}$
would give an $SU(2)$ doublet with
$Y= -1 $ for $p =q$ and $Y = 1$ for $q = p+3$.
Thus for a vector field, we need three types of representations:
\begin{enumerate}
\item $R_1\equiv (p,p)$-type, $p \neq 0$: These contribute to both $A_i$ and $A_{\bar i}$
\item $R_2 \equiv (p, p+3)$-type: These contribute to $A_i$.
\item $R_3 \equiv (p+3, p)$-type: These contribute to $A_{\bar i}$
\end{enumerate}
The general expression for an Abelian vector field is thus
\beqar
A_i &=& \sum_{A, R_1} C^{R_1}_A \, \la R_1, A\vert {\hat g} \vert R_1, i\ra
+ \sum_{A, R_3} B^{R_3}_A \, \la R_3, A\vert {\hat g} \vert R_3 , i\ra
\nonumber\\
A_{\bar j}  &=& \sum_{A, R_1} {\bar C}^{R_1}_A \, \la R_1, A\vert {\hat g} \vert R_1, {\bar j} \ra
+ \sum_{A, R_2} {\bar B}^{R_2}_A \, \la R_2, A\vert {\hat g} \vert R_2 , {\bar j}\ra
\label{4}
\eeqar
where $C^{R_1}_A$ and $B^{R_3}_A$ are arbitrary complex numbers.
The representations $R_2$ and $R_3$ are conjugates of each other; $R_1$ is 
invariant under conjugation. The state on the right for $R_1$, namely,
$T^{i 3 \cdots 3}_{3\cdots 3}$ can be obtained by the action of
$t_{4}+ i t_5$ and $t_6 + i t_7$ on a state $\vert w\ra$ of the form
$T^{3\cdots 3}_{3\cdots 3}$, which is $SU(2)$ invariant with zero hypercharge. 
In other words it can be obtained by the action of
$\nabla_i$ on a function. Thus the first terms in (\ref{4}) are of the form of derivatives acting
on a function. In a similar way, the relevant state $\vert R_3, i\ra$ can be obtained by the
action of $\epsilon _{ij} \nabla_{\bar j}$ on a state $\vert z\ra$ which is $SU(2)$ invariant with
$Y= 2$. (A point of clarification: Even though there is only one irreducible doublet representation for $SU(2)$, it is only pseudo-real. Thus to convert doublet with upper indices to ones with lower indices, we have to use the $\epsilon_{ij}$ symbol.) In a similar way, we can obtain the
relevant state for $\vert R_2, {\bar j}\ra$ by the action of
$\epsilon_{ij} \nabla_j$ on a state which is $SU(2)$ invariant with $Y = -2$.
Combining these results, we see that the parametrization given above reduces to
\beqar
A_i &=&  - \nabla_i \theta  + \epsilon_{ij} \, \nabla_{\bar j} \chi
\nonumber\\
A_{\bar j}  &=&  \nabla_{\bar i} {\bar \theta } - \epsilon_{ij} \, \nabla_{ j} {\bar\chi}
\label{5}
\eeqar
(In preparation for the nonabelian case where we use antihermitian basis for the gauge fields,
we have changed over, compared to (\ref{4}), to the conjugation property $A_i^\dagger = - A_{\bar i}$. In other words,
$A_i$ correspond to $-i( A_4+i A_5), -i (A_6 +i A_7)$.)
In (\ref{5}), $\vf$ is a complex function on $\mathbb{CP}^2$ and hence is expandable in terms
of $\la R_1, A\vert {\hat g}\vert w\ra$. 
The quantity $\chi$ is expandable in terms
of $\la R_3, A\vert {\hat g}\vert z\ra$; it does not define a function on $\mathbb{CP}^2$ since
$\vert z\ra$ has $Y= 2$. The term $\epsilon_{ij} \nabla_{\bar j} \chi$ may be thought of
as the divergence of a two-form.
The four real independent components for a general vector field in four dimensions
are captured by the $\theta$, ${\bar \theta}$, $\chi$ and ${\bar \chi}$.

The generalization to the nonabelian case is straightforward. Notice that, the product of
a state of the form $\vert w\ra$ with another state of the form $\vert w\ra$ still gives a state of the same type. Thus functions can be multiplied to form other functions.
Also, the product of a state of the type $\vert w\ra$ with
$\vert z\ra$ still gives a state of the form
$\vert z\ra$. Thus multiplying $\chi$ by functions is also possible.
We may combine this and write a parametrization for $A_i$ as
\beq
A_i = - \nabla_i M \, M^{-1} - M \, a_i \, M^{-1}
\label{6}
\eeq
where $M$ is a complex matrix in the complexification of the gauge group. We will take the gauge group as $SU(N)$ for the rest of this paper, so $M \in SL(N, {\mathbb C})$.
Gauge transformations act on $M$ as $M \rightarrow M^U = U \, M$; the term
$M a_i M^{-1}$ transforms covariantly under this, so that $A_i$ in
(\ref{6}) has the expected transformation property
\beq
A_i \rightarrow A^U_i = U \, A_i \, U^{-1} - \nabla_i U \, U^{-1}
\label{7}
\eeq
The conjugate components are given by
\beq
A_{\bar i} = M^{\dagger -1} \nabla_{\bar i} M^\dagger + M^{\dagger -1} {\bar a}_{\bar i}\,
M^\dagger
\label{8}
\eeq
Since the inhomogeneous parts in the gauge transformation are generated from
$-\nabla_i M \, M^{-1}$ and $M^{\dagger -1} \nabla_{\bar i} M^\dagger$,
$D_i f \equiv \nabla_i f+ [ - \nabla_i M \, M^{-1}, f]$ and
${\bar D}_{\bar i} \equiv \nabla_{\bar i} f + [M^{\dagger -1} \nabla_{\bar i} M^\dagger , f]$
are gauge-covariant derivatives, for $f$'s which transform as
$f \rightarrow f^U = U f U^{-1}$.
Thus another way to generalize (\ref{5}) is
\beqar
A_i = - \nabla_i M \, M^{-1} + \epsilon_{ij} {\bar D}_{\bar j} \phi\nonumber\\
A_{\bar i} = M^{\dagger -1} \nabla_{\bar i} M^\dagger - \epsilon_{ij} D_j \phi^\dagger
\label{9}
\eeqar
Here $\phi$ transforms covariantly under gauge transformations.
Primarily, the parametrization of the gauge fields we use will be
(\ref{9}). But we may also view it as equivalent to (\ref{6}, \ref{8}), defining
\beq
a_i = - M^{-1} \epsilon_{ij} {\bar D}_{\bar j} \phi \, M , \hskip .3in
{\bar a}_{\bar i} = - M^\dagger \, \epsilon_{ij} D_j \phi^\dagger \, M^{\dagger -1}
\label{10}
\eeq
Both these ways of viewing the parametrization of the gauge fields
will be useful later. In terms of the matrix structure, the gauge fields are of the form
$A_1 = (-iT^a) A^a_1= (-i T^a) \, (A^a_4 + i A^a_5)$, $A_{\bar 1} = (-iT^a) A^a_{\bar 1} = (-i T^a) \, (A^a_4 - i A^a_5)$, etc.,
where $\{ T^a \}$ form a basis for the Lie algebra of the gauge group, say, $SU(N)$.

The gauge transformation properties show that the gauge-invariant degrees of freedom are described by $H = M^\dagger M$ and $\chi = M^{-1} \phi \, M$,
${\bar \chi} = M^\dagger \phi^\dagger \, M^{\dagger -1}$ (or $M^{ba} \phi^b$ and $(M^\dagger)^{ab} \phi^{\dagger b}$).  (We use the same letter $\chi$, although these are matrices and
parametrize the nonabelian 
fields now.)
These fields constitute the coordinates for the space of gauge-invariant
configurations, i.e., coordinates for the gauge-orbit space $\C$.

\subsection{The metric and volume}

We now turn to the metric on the space of these gauge potentials. It is given by
\beq
ds^2 = - 2 \int d \mu \, \Tr (\delta A_{\bar i} \, \delta A_i) = \int d\mu~ \delta A^a_{\bar i}\,
\delta A^a_i
\label{11}
\eeq
where $d\mu$ is the volume element for $\mathbb{CP}^2$.
Taking the variations of (\ref{9}) we find
\beqar
\delta A^a_i &=& - (D_i \theta)^a + \epsilon_{ij} ({\bar D}_{\bar j} \delta \phi )^a
+ \epsilon_{ij} f^{abc} ({\bar D}_{\bar j} \theta^\dagger )^b \, \phi^c \nonumber\\
\delta A^a_{\bar i} &=&  -({\bar D}_{\bar i} \theta^\dagger)^a + \epsilon_{ij} ({D}_{ j} \delta \phi^\dagger)^a
+ \epsilon_{ij} f^{abc}  ({D}_{j} \theta )^b \, \phi^{\dagger c}
\label{12}
\eeqar
where $\theta = \delta M \, M^{-1} = (-i T^a ) \theta^a$ and
$f^{abc}$ are the structure constants of the Lie algebra defined by
$[T^a, T^b] = i f^{abc} T^c$.
Using these variations in (\ref{11}), we obtain
\begin{align}
ds^2 &= (ds^2)_0 + (ds^2)_1  + (ds^2)_2\nonumber\\
(ds^2)_0 &=   \int d \mu ~ \left[ (  {\bar D}_{\bar i} \theta^\dagger - \epsilon_{ij} {D}_{ j} \delta \phi^\dagger)^a (D_i \theta - \epsilon_{ik} {\bar D}_{\bar k} \delta \phi )^a\right]
\nonumber\\
(ds^2)_1&= \int d \mu ~ \left[  - \epsilon_{ij} f^{abc} ({D}_{j} \theta )^b \, \phi^{\dagger c}
(D_i \theta - \epsilon_{ik} {\bar D}_{\bar k} \delta \phi )^a
- (  {\bar D}_{\bar i} \theta^\dagger - \epsilon_{ij} {D}_{ j} \delta \phi^\dagger)^a
 \epsilon_{ik} f^{abc}({\bar D}_{\bar k} \theta^\dagger )^b \, \phi^c \right]\nonumber\\
 (ds^2)_2&=\int d\mu~ \left[   \epsilon_{ij} \epsilon_{ik} f^{abc} f^{amn}({D}_{j} \theta )^b \, \phi^{\dagger  c} \, ({\bar D}_{\bar k} \theta^\dagger )^m \, \phi^n\right]\label{13}
\end{align}
We have separated the metric into terms with no power of $\phi$ or $\phi^\dagger$,
with one power of the same, or two powers. It is worth emphasizing that the connections in the covariant derivatives $D_i$ and $\bD_{\bar i}$ are $- \nabla_i M \, M^{-1}$ and
$M^{\dagger -1} \nabla_{\bar i} M^\dagger$, respectively. As a result, we
can further simplify $(ds^2)_0$ by partial integration, by noting that
\beqar
\int d\mu \left[ \epsilon_{ij} ({D}_{ j} \delta \phi^\dagger)^a \, (D_i \theta)^a\right]
&=& - \int d\mu \left[ \delta \phi^{\dagger a}\,  \epsilon_{ij}  (D_j D_i \theta)^a\right]\nonumber\\
&=& 0
\label{14}
\eeqar
Since $D_i$ only involves $\nabla_i M \, M^{-1}$, the holomorphic covariant derivatives commute
and so $\epsilon_{ij} D_i D_j =0$.
Thus
\beq
(ds^2)_0 = \int d \mu ~ \left[ \theta^{\dagger a} ( - {\bar D}_{\bar i} \, D_i )^{ab}  \theta^b
+ \delta \phi^{\dagger a}( - D_i \, {\bar D}_{\bar i} )^{ab} \delta \phi^b\right]
\label{15}
\eeq
We can simplify the other terms in $ds^2$ in a similar way to get
\beqar
(ds^2)_1 &=& \int d\mu~ \left[ \theta^a ( \epsilon_{ik} D_i \, \Phi^\dagger  \, D_k )^{ab} \theta^b
+ \theta^a ( D_k\, \Phi^\dagger \, {\bar D}_{\bar k})^{ab} \delta \phi^b \right.\nonumber\\
&&\hskip .5in \left. 
+ \theta^{\dagger a} ( \epsilon_{ik} {\bar D}_{\bar i}\, \Phi\, {\bar D}_{\bar k} )^{ab} \theta^{\dagger b}
+ \delta\phi^{\dagger a} ( - D_k \, \Phi \,  {\bar D}_{\bar k} )^{ab} \theta^{\dagger b}\right]
\nonumber\\
(ds^2)_2 &=& \int d\mu~ \left[  \theta^{\dagger a} ( {\bar D}_{\bar k} \, \Phi \, \Phi^\dagger \, D_k )^{ab}
\theta^b\right]
\label{16}
\eeqar
where $\Phi $ is a matrix, $(\Phi)^{ab} = \phi^c \, f^{abc}$ and
$(\Phi^\dagger)^{ab} = \phi^{\dagger c} f^{abc}$. Define a $4\times 4$ matrix of operators 
$\M$ by
\begin{align}
\M_{11} &= \M_{33} =  (-{\bar D}_{\bar i} D_i \, + {\bar D}_{\bar k} \, \Phi \, \Phi^\dagger\, D_k)
\nonumber\\
\M_{22}&= \M_{44} = (- D_i {\bar D}_{\bar i} )\nonumber\\
\M_{13} &= 2 \,(\epsilon_{ik} {\bar D}_{\bar i} \, \Phi \, {\bar D}_{\bar k} ), \hskip .3in
\M_{23} = 2\,  (- D_k \Phi {\bar D}_{\bar k} )\nonumber\\
\M_{31} &= 2\,  (\epsilon_{ik} D_i \, \Phi^\dagger \, D_k ), \hskip .3in
\M_{32} = 2 \,(D_k \Phi^\dagger \, {\bar D}_{\bar k})
\label{17}
\end{align}
with all other elements being zero. Then the metric is
\beq
ds^2 = {1\over 2} \int d\mu ~ \xi^\dagger_A \, \M_{AB} \, \xi_B , \hskip .3in
(\xi_1, \xi_2 , \xi_3 , \xi_4 ) = ( \theta, ~ \delta \phi , ~ \theta^\dagger ,~ 
\delta\phi^\dagger )
\label{18}
\eeq
The volume element corresponding to this is given, up to an overall normalization factor,
 by $\sqrt{\det \M }$ times the volume defined by the differentials $\xi$.
 For the latter, $\theta$ and $\theta^\dagger$ give the standard Cartan-Killing
 volume element of
 $SL(N, {\mathbb C})$ (at each spacetime point); the differentials 
 $\delta \phi$, $\delta \phi^\dagger$ give the standard functional integration measure
 $[d \phi\, d\phi^\dagger]$. Thus the volume element corresponding to
 (\ref{18}) becomes
 \beq
 dV = \sqrt{\det \M} ~ d\mu_{SL(N, {\mathbb C})} ~ [d \phi \, d\phi^\dagger]
 \label{19}
 \eeq
 (In $d\mu_{SL(N, {\mathbb C})} $, we have a product of the volume of 
 $SL(N, {\mathbb C})$ over all spacetime points; this is not explicitly displayed, but
  left as understood.)
 As discussed in \cite{KKN}, by doing a polar decomposition of $M$ into a unitary matrix and a hermitian matrix, we can factor out the volume of gauge transformations from $d\mu_{SL(N, {\mathbb C})}$,
 \beq
 d\mu_{SL(N, {\mathbb C})} =  d\mu (H)~ d\mu(SU(N))
 \label{20}
 \eeq
 The integration measure for the $\phi$'s is gauge-invariant since these fields transform 
 covariantly, in much the same way matter fields have a gauge-invariant measure
 in standard functional integration.
 Factoring out the volume of gauge transformations, we get the volume element for the gauge-orbit space $\C$ (or the space of physical configurations) as
 \beq
 d\mu (\C )  = \sqrt{\det \M} ~ d\mu (H) ~ [d \phi \, d\phi^\dagger]
 \label{21}
 \eeq
 Also, as noted in \cite{KKN},
parametrizing the matrix $H$ in terms of a set of real fields $\lambda^a$, we can write
 $H^{-1}dH = d\lambda ^a\,  r_{ak} (\lambda)\, T^k$ and
 $d \mu (H) = \prod_x (\det r) \, [d\lambda ]$.
 
 \subsection{Calculating the Jacobian factor}
 
The problem is now reduced to the computation of the determinant of $\M$. 
For this we note that the off-diagonal terms in the matrix $\M$ depend on
$\Phi$ or $\Phi^\dagger$, so in the neighborhood of the subspace
$\Phi = 0$, $\M$ has only diagonal elements given by the operators
$(- {\bD}_{\bar k} D_k)$ and $(- D_k \bD_{\bar k})$.
Our strategy will be to calculate the volume around this subspace.
The terms which depend on $\Phi$, $\Phi^\dagger$ can then be included in a series expansion.
The gauge transformation of the potentials $A_i$, $A_{\bar i}$ is fully captured by $M$
and $M^\dagger$ in the parametrization we have used; thus setting
$\Phi$ and $\Phi^\dagger$ to zero (in $\sqrt{\det \M}$) is consistent with gauge invariance requirements. Taking a Hamiltonian point of view for a moment, the
two polarization states which would normally be eliminated by the Gauss law
-which being a first class constraint eliminates two degrees of freedom- are contained in
$M$, $M^\dagger$. The fields
 $\Phi$ and $\Phi^\dagger$ act almost as matter fields describing the surviving two polarizations.
 We may therefore expect to gain some insight into many of the issues of low energy physics 
 from the analysis of the measure near the subspace with $\Phi = \Phi^\dagger = 0$.
 
 The quantity to be calculated is thus $\log [\det ( - \bD_{\bar k} \, D_k ) \, \det ( - D_k\,\bD_{\bar k} )] $.
In two dimensions, the analogous quantity would be $\log \det (- \bD \, D )$.
Formally we can factorize this, calculating $\log \det \bD$ and $\log \det D$ separately and putting them together with the standard Schwinger-Quillen counterterm to obtain the gauge invariant result.
In four dimensions, such a factorization is obviously not possible since we have a sum over the two complex indices in $\bD_{\bar k} D_k$.
So we will first recalculate the two-dimensional case in a way that will help us generalize
to the four dimensions.
In two dimensions, we need to calculate $\Gamma = \log \det (- \bD \, D ) =
\Tr \log ( - \bD \, D )$. Consider the variation of this with respect to $A$. We  get
\beq
\delta \Gamma = \Tr \left[ - \delta (\bD D ) \, \left( {1\over - \bD D }\right)\right]
= \int d^2x ~\Tr \left[ - \bD_x ( \delta A_x \, G(x, y) )\right]_{y \rightarrow x}
\label{22}
\eeq
where $G (x, y) = (-\bD D )^{-1}_{x,y}$. The operator $\bD$ acts on both $\delta A$ and
$G(x,y)$. When it acts on $\delta A$, we have $G(x,y)\bigr]_{y \rightarrow x}$.
This is proportional to the identity in any regularized version of $G$ and hence this contribution vanishes by the matrix trace. The surviving term is
\beq
\delta \Gamma
= \int d^2x ~\Tr \left[ - \delta A_x \,\bD_x G(x, y) \right]_{y \rightarrow x}
\label{23}
\eeq
(We have written out the functional trace; the remaining trace is just over the matrices.)
This shows that we need a regularized version of the short-distance behavior of $\bD_x G(x,y)$.
We know that $(-\bdel \del )^{-1}$ behaves as $\log [ (x-y)(\bx - \by)]$ at short separations, so that
$\bdel G(x,y)$ behaves as $1/(\bx - \by )$.  However, we need to put in phase factors which ensure the correct gauge transformation properties. It can then be seen that the short-distance behavior should be given by
\beqar
\bD_x G(x,y) &\simeq& - {M(x) \, M^{-1} (y) \, W (y,x) \over
\pi ( \bx - \by ) }\nonumber\\
W(y,x) &=& P \exp (-\int^y_x A )
\label{24}
\eeqar
$W(y,x)$ is the Wilson line matrix which transforms under gauge transformations as
\beq
W(y, x) \rightarrow W(y,x, A^U) = U(y) \, W(y,x) \, U^{-1}(x)
\label{25}
\eeq
so that $\bD_x G(x,y) \rightarrow U(x) \, \bD_x G(x,y) \, U^{-1} (x)$. Since
$\delta A$ transforms covariantly, $\delta A_x \rightarrow  U (x) \, \delta A_x \, U^{-1} (x)$,
this makes the trace in (\ref{23}) gauge invariant. Further, since the numerator
of $\bD_x G(x,y)$ in (\ref{24}) transforms covariantly,
we have 
\beqar
D (M(x) \, M^{-1} (y) \, W (y,x) ) &=&
\del (M(x) \, M^{-1} (y) \, W (y,x)) 
+ [ A\, , M(x) \, M^{-1} (y) \, W (y,x)\, ] \nonumber\\
&=& 0
\label{26}
\eeqar
The action of $\del$ on $1/\pi(\bx - \by )$ leads to a delta function, verifying
\beq
- D_x [ \bD_x G(x,y) ] = \delta^{(2)}(x-y)
\label{27}
\eeq
This verifies the correctness of the short-distance behavior of the $\bD_x G(x,y)$ given in
(\ref{24}).

It is now straightforward to expand (\ref{24}) to first order in $x-y$, $\bx-\by$ and find
\beqar
\delta \Gamma &=& {1\over \pi} \int d^2x~ \Tr [ \delta A ( \bA + \bdel M \, M^{-1} ) ]\nonumber\\
&=& {1\over \pi} \int d^2x~ \left[ \Tr (\delta A \, \bA ) - \Tr (\bdel \theta \, \del M \, M^{-1} )\right]
\nonumber\\
&=& {1\over \pi} \int d^2x~ \Tr (\delta A \, \bA ) + \delta S_{wzw} (M)
\label{28}
\eeqar
where the WZW action is given by
\beq
S_{wzw} (M) = {1\over 2\pi} \int d^2x~ \Tr ( \del M \, \bdel M^{-1} )
+{i \over 12 \pi} \int \Tr ( M^{-1} d M )^3
\label{29}
\eeq
There is a similar result for the variation of $M^\dagger$ or $\bA$ and the combined result is
\beq
\Gamma = S_{wzw} (H)
\label{30}
\eeq
In arriving at (\ref{28}), we have used the symmetric way of taking the limit $y \rightarrow x$, so that
$(x-y)/(\bx - \by )$ gives zero.

Before going to the four-dimensional case, there is one other point worth emphasizing.
In parametrizing the fields as $A = - \del M \, M^{-1}$,
$\bA = M^{\dagger -1} \bdel M^\dagger$ there is an ambiguity
since $M$ and $M V (\bx )$, where $V(\bx )$ is antiholomorphic,  give the same
$A$.
The use of $M$'s in (\ref{24}) carry this ambiguity over to the short-distance behavior.
However, it is immaterial, as the corresponding correction to
$\delta \Gamma$ vanishes,
\beqar
\delta \Gamma \,\bigr]_{M\rightarrow MV}&=&   {1\over \pi} \int \Tr [ \delta A M \, \bdel V \, V^{-1} \, M^{-1}]
= {1\over \pi} \int \Tr [ - D \theta  M \, \bdel V \, V^{-1} \, M^{-1}]\nonumber\\
&=&{1\over \pi} \int \Tr [ \theta \, D (M \bdel V \, V^{-1} \, M^{-1}) ]
= {1\over \pi} \int \Tr [ \theta M \, \del (\bdel V \, V^{-1}) \, M^{-1}]\nonumber\\
&=& 0\label{31}
\eeqar

In four dimensions, the connections in the covariant derivatives are $-\nabla_k M \, M^{-1}$
and $M^{\dagger -1} \nabla_{\bar k} M^\dagger$. The degree of divergence for the short-distance behavior is worse since
$ G(x,y) \sim (- \nabla_{\bar k} \nabla_k )^{-1}_{x,y} \sim (x -y )^{-2}$. We will introduce
a Pauli-Villars type regulator which corresponds to the replacement
\beq
\left( {1\over - \bD_{\bar k} \, D_k }\right) \rightarrow 
\left( {1\over - \bD_{\bar k} \, D_k }\right) \left( {\Lambda^2 \over - \bD_{\bar j} D_j + \Lambda^2}\right)
\equiv G_{reg}(x,y)
\label{32}
\eeq
The parameter $\Lambda^2$ (with the dimension of (mass)$^2$) is the ultraviolet cut-off.
The short-distance behavior of this function is given by
$G_{reg}(x, y) \sim \Lambda^2 (-\bdel \del )^{-2} \sim \Lambda^2 \log (x-y)^2$, just as in two dimensions.
Thus we get the short-distance behavior
\beq
\bD_{\bar k} G (x,y)\, \bigr]_{reg} \simeq - {M^2 \over \pi} {(x -y )_k \over \vert x - y\vert^2}\,
(M(x) \, M^{-1} (y) \, W (y,x) ) 
\label{33}
\eeq
(The numerical factors are not quite precise; it is immaterial 
since they can all be absorbed into
$M^2$.)
As before, defining $\Gamma = \Tr \log (- \bD_{\bar k} D_k)$ we find
\beqar
\delta \Gamma &=& \int d \mu ~ \Tr \left[ \delta (- \nabla_k M \, M^{-1}) \,( - \bD_{\bar k} G(x,y))_{reg}
\right]_{y \rightarrow x}\nonumber\\
&=& {\Lambda^2 \over \pi} \int d\mu ~ \Tr \left[ \delta (- \nabla_k M \, M^{-1}) \,(M^{\dagger -1} \nabla_{\bar k}
M^\dagger  + \nabla_{\bar k} M\, M^{-1})
\right]
\label{34}
\eeqar
We have taken the angular symmetric limit as $y \rightarrow x$, so that
\beq
{(x-y)_k \, (\bx -\by )_{\bar a} \over \vert x - y \vert^2}\bigr]_{y \rightarrow x} =
c \, \delta_{ka}
\label{35}
\eeq
for some constant $c$, which has been absorbed into the cut-off $M^2$.
We have a similar result for the variation with respect to $M^\dagger$ and the results can be combined to obtain
\beq
\Gamma = \Lambda^2 \, S_{4d} (H)
\label{36}
\eeq
where $S_{4d}$ is the four-dimensional WZW action appropriate to a four-dimensional
K\"ahler manifold.
This action is basically contained in Donaldson's paper \cite{Don}, but was independently derived as the boundary action for the K\"ahler-Chern-Simons theory in \cite{NS} in an attempt to generalize conformal field theories to four dimensions. It has since been studied by a number of authors, most notably starting with the work of Losev {\it et al} \cite{Los}.
For an arbitary matrix $N$, it is explicitly given by
\beqar
S_{4d} (N)  &=& {1\over 2 \pi} \int d\mu ~ \Tr ( \nabla_k N \, \nabla_{\bar k} N^{-1} )
+ {i \over 12 \pi} \int \omega\wedge \Tr (N^{-1} d N )^3\nonumber\\
&=&  {1\over 2 \pi} \int d\mu ~ g^{a{\bar a}}\,\Tr ( \del_a N \, \bdel_{\bar a} N^{-1} )
+ {i \over 12 \pi} \int \omega\wedge \Tr (N^{-1} d N )^3\label{37}
\eeqar
where $\omega$ is the K\"ahler form for $\mathbb{CP}^2$. 
For $\Gamma$, we need $S_{4d} (M^\dagger M ) = S_{4d} (H)$.
In the first term of the first line of (\ref{37}), we are still using the derivatives in the tangent frame (given by the right translation operators on the group element which coordinatizes the manifold). In the second line, we 
show the expression in terms of the derivatives in the local coordinate description, with
$g^{a {\bar a}}$ as the inverse to the K\"ahler metric $g_{a {\bar a}}$. In local coordinates, the metric and the K\"ahler form are given by
\beqar
ds^2 &=& \left[ {d\bz \cdot dz 
\over (1+\bz \cdot z) }- {z \cdot d\bz ~\bz\cdot dz \over
(1+\bz \cdot z )^2}\right] \equiv g_{a {\bar a}} dz^a \, d\bz^{\bar a}\nonumber\\ 
\omega &=& {i \over 2} g_{a {\bar a}} dz^a \wedge d\bz^{\bar a}
\label{38}
\eeqar
In this convention,
\beq
d\mu = {1\over 4} ( \det g_{a{\bar a}}) \, dz^1 d\bz^{\bar 1} dz^2 d\bz^{\bar 2}
= (\det g ) \, d^4x
\label{39}
\eeq
This higher dimensional WZW action also satisfies a Polyakov-Wiegmann identity of the form
\cite{{PW}, {NS}}
\beq
S_{4d} (N M) = S_{4d} (N) + S_{4d} (M) -
{1\over \pi}  \int d\mu~ \Tr ( N^{-1} \nabla_{\bar a} N \, \nabla_a M \, M^{-1} )
\label{40}
\eeq
This is easily verified by direct substitution and simplification in
(\ref{37}). This identity shows that $\Gamma$ in (\ref{36}) satisfies
(\ref{34}), thereby justifying (\ref{36}) as the integrated version of (\ref{34}).

There are a number of refinements to be considered.
First of all, so far we have only calculated
 the leading term proportional to $\Lambda^2$; there can be subleading terms and finite terms, which are not captured by (\ref{36}) because of the way we have taken the
short distance limit. So the result (\ref{36}) should, more accurately,  be expressed as
\beq
\Tr \log (- \bD_{\bar k} D_k ) =  \Lambda^2 \, S_{4d} (H) ~+~ {\rm subleading ~+~ finite~ terms}
\label{41}
\eeq
Secondly, for the measure calculation, we also need $\det ( - D_k \bD_{\bar k})$. Notice that
if we make the transformation $z^a \leftrightarrow \bz^{\bar a}$ and
$\nabla_k \leftrightarrow \nabla_{\bar k} $ and $M \leftrightarrow M^{\dagger -1}$, then $D_k
\leftrightarrow \bD_{\bar k}$. So this second determinant is the same as the first with
$z^a \leftrightarrow \bz^{\bar a}$ and $H \leftrightarrow H^{-1}$. The first term of $S_{4d}$ is obviously unchanged; the second term changes sign under $H \leftrightarrow H^{-1}$ and there is another minus sign from $z^a \leftrightarrow \bz^{\bar a}$. So it is unchanged as well
as we find
\beq
\Tr \log ( - D_k \bD_{\bar k} ) = \Lambda^2 S_{4d} (H) ~+~ {\rm subleading ~+~ finite~ terms}
\label{42}
\eeq
Going back to (\ref{21}), we can now write our result so far as
\beqar
d\mu (\C ) &=& \sqrt{\det \M} ~ d\mu (H) ~ [d \phi \, d\phi^\dagger]\nonumber\\
&\approx& \det (- \bD_{\bar k} D_k ) \, \det (- D_m \bD_{\bar m} ) ~ d\mu (H) ~ [d \phi \, d\phi^\dagger]
\nonumber\\
&\approx& e^{ 2 \Lambda^2 \, S_{4d} (H) } ~ d\mu (H) ~ [d \phi \, d\phi^\dagger]
\label{43}
\eeqar

We will now look at how this result can be improved by some of the $\Phi , \Phi^\dagger$-dependent terms.
Separating off the $\M_{13}, \, \M_{23}$, etc., we can write
$ \log \sqrt{\det \M }$ as
\begin{align}
\log \sqrt{\det \M} &= {1\over 2} \Tr \log \M \nonumber\\
&= \Tr \log \M_{11} + \Tr \log \M_{22} 
+ {1\over 2} \Tr \log ( 1 + {\bf X})\nonumber\\
&= \Tr \log (- \bD_{\bar k} D_k  + \bD_{\bar k}\, \Phi\, \Phi^\dagger D_k) + \Tr \log (-D_m \bD_{\bar m} )
+{1\over 2}  \Tr {\bf X} - {1\over 4} \Tr ( {\bf X} {\bf X} ) + \cdots\nonumber\\
{\bf X}&= \left[ \begin{matrix}
0& 0& \M_{11}^{-1} \M_{13}\\
0&0&\M_{22}^{-1} \M_{23}\\
\M_{11}^{-1} \M_{31} & \M_{11}^{-1} \M_{32}&0\\
\end{matrix} 
\right]
\label{44}
\end{align}
The term in $\log\sqrt{\det\M}$ which is second order in $\Phi$, $\Phi^\dagger$ is then
\begin{align}
(\log\sqrt{\det \M})_2 &=
\int d\mu_x \Tr \left[ \bD_{\bar k}  \Phi \Phi^\dagger D_k G(x,y) \right]_{y\rightarrow x}\nonumber\\
& \hskip .2in - 2 \int d\mu_x d\mu_y\,
\epsilon_{ik} \epsilon_{mn}
\Tr \left[ D_i \Phi^\dagger D_k G(x,y) \bD_{\bar m} \Phi \bD_{\bar n} G(y,x)\right]\nonumber\\
& \hskip .2in + 2 \int d\mu_x d\mu_y\,
\Tr \left[ D_k \Phi\, \bD_{\bar k}  G(x,y) D_{m} \Phi^\dagger \bD_{\bar m} {\widetilde G(y,x)}\right]
\label{45}\\
G(x,y)&= \left( - \bD_{\bar k} D_k \right)^{-1}_{x,y}, \hskip .3in
{\widetilde G(x,y)} = \left(- D_k \bD_{\bar k} \right)^{-1}_{x,y}
\label{46}
\end{align}
This term can be calculated with a suitable regulator and clearly one can go to higher powers
as there is a systematic expansion of
(\ref{44}) in powers of
 $\Phi$, $\Phi^\dagger$. Nevertheless, it is an involved process and we will postpone further discussion of this. It will not be needed for the arguments presented in the next section.
 Instead, what
we will do here is to determine some of the quadratic terms in $\Phi$, $\Phi^\dagger$ using a symmetry argument. For this we go back to the parametrization (\ref{6}) and (\ref{8}).
Notice that the components of $a_i$ and $\ba_{\bar i}$ are all related to the single
complex quantity $\phi$, as in (\ref{10}). 
However,
consider evaluating the Jacobian factor for the $\theta$, $\theta^\dagger$ part of
$d\mu (\C )$ for arbitrary
$a_i$ and $\ba_{\bar i}$, setting them to the values given in (\ref{10}) at the end.
This can be done by taking variations of (\ref{6}) and (\ref{8}) at fixed $a_i$, $\ba_{\bar i}$.
Notice that (\ref{6}) and (\ref{8}) have something of a ``fake gauge symmetry",
\begin{align}
M \rightarrow M^S = M\, S , \hskip .5in a_i \rightarrow ~&a^S_i = S^{-1} a_i S - S^{-1} \nabla_i S 
\nonumber\\
M^\dagger \rightarrow M^{\dagger S} = S^{-1} M^\dagger,
\hskip .3in
\ba_{\bar i} \rightarrow ~&\ba_{\bar i}^S = S^{-1} \ba_{\bar i} S + S^{-1} \nabla_{\bar i} S
\label{47}
\end{align}
The calculation of the Jacobian must have this symmetry implying that we must consider the gauged version of the WZW action. This is given by
\beq
S_{4d} (H, a, \ba) = 
S_{4d}(H) - {1\over \pi} 
\int d\mu~ \Tr \left[ H^{-1} \nabla_{\bar i} H\, a_i + \ba_{\bar i} \nabla_i H \, H^{-1}
+ \ba_{\bar i}\, H \, a_i \, H^{-1} - \ba_{\bar i} \, a_i\right]
\label{48}
\eeq
We can now substitute for $a_i$, $\ba_{\bar i}$ from (\ref{10}) and simplify to get
 \begin{align}
 S_{4d}(H, \chi, {\bar\chi})=
 S_{4d} (H) - {1\over \pi} \int d \mu ~ &\Tr \left[ H^{-1} ({\cal D}_a{\bar\chi} ) \, H \,{\cal D}_a \chi
 - ({\cal D}_a{\bar\chi} ) \, {\cal D}_a \chi\right.\nonumber\\
&\hskip.2in \left.  - \epsilon^{\ba {\bar b} } 
 H^{-1} \nabla_{\bar a} H \, {\cal D}_{\bar b} \chi
 - \epsilon^{ab} {\cal D}_b {\bar \chi} \, \nabla_a H \, H^{-1}
 \right]
 \label{49}
 \end{align}
 where, as stated before, $\chi = M^{-1} \phi M$, ${\bar \chi} = M^\dagger \phi^\dagger M^{\dagger -1}$. Since we can consider $[d\phi \, d\phi^\dagger] = [d\chi \, d{\bar \chi}]$ as well,
 we can now summarize our results so far as follows. 
 \beq
 d\mu (\C ) = d\mu (H) \, [d\chi \, d{\bar \chi}] \, \exp \left[ 2 \Lambda^2 \, S_{4d}(H, \chi, {\bar\chi})
+ \dots \right]
\label{50}
\eeq
where the ellipsis denotes terms which are subleading in the divergence, or finite, or involve higher powers of $\chi$, $\bar\chi$.

\section{Discussion}
We are now in a position to discuss the relevance of this result (\ref{50}) for the functional integration in a gauge theory.

First of all, note that the term $S_{4d}$ is only obtained for the nonabelian theory. The variables 
$\theta^a$ and $\phi^a$ transform in the adjoint representation and 
the trace in $S_{4d}$ is in the same representation, hence vanishing for the Abelian theory.
Secondly, we note that $S_{4d}$ has the properties of a mass term for the gauge fields in the sense of being a gauge-invariant completion of $A_i \bA_{\bar i}$. In fact, it is well known that the WZW action in two dimensions is a mass term for the gauge fields \cite{PW}; this even goes back to Schwinger's original calculation in the Abelian case. It is also known that such a term defined on a lightcone (with a suitable integration over the orientations of the lightcone) can describe the screening mass in four-dimensional Yang-Mills theory at finite temperature \cite{HTL}.

Usually, when we integrate over fermions in a four-dimensional gauge theory, there is a quadratic divergence proportional to $A^2$, but this is generally rejected on the grounds that there is no such term which is both gauge and Lorentz invariant. In other words, there is no such term consistent with gauge invariance and the isometries of the underlying space.
(And, indeed, with a gauge- and Lorentz-invariant regularization, no such term is generated.)
However, in our case, the term we find is gauge invariant and invariant under the isometries of the space ${\mathbb{CP}^2}$.
Therefore the conclusion is that we must define the gauge theory by including such a term from the beginning with a bare parameter $m^2_0$, so that
\beqar
d\mu (\C ) &=& d\mu (H)\, [d\chi d{\bar \chi}]\, \,\sqrt{\det \M}\,\, \exp \left[ m^2_0 \, S_{4d} (H, \chi, {\bar \chi})\right]
\nonumber\\
&=& d\mu (H)\, [d\chi d{\bar \chi}]\, \,\exp \left[ m^2_R \, S_{4d} (H, \chi, {\bar \chi}) + \dots\right]
\label{51}
\eeqar
The renormalized value of this parameter, namely $m^2_R$, then defines a mass scale for the theory.
Thus the functional integral for Yang-Mills theory would be defined as
\beqar
Z &=& \int d\mu (\C ) \, e^{- S_{YM}(A)}\nonumber\\
&=& \int d\mu (H) [d\chi d{\bar \chi}]\, \sqrt{\det \M}\, \exp \left[ m^2_0 \, S_{4d} (H, \chi, {\bar \chi})\right]\,
e^{- S_{YM}(H, \chi, {\bar \chi})}
\label{52}
\eeqar
This is the main conclusion of this paper.
For the term $ \exp \left[ m^2_0 \, S_{4d} (H, \chi, {\bar \chi})\right]$ which we need to have for a well-defined definition of the integration measure, it is sufficient to understand the
divergence structure of $\sqrt{\det \M}$, which can then determine the
nature of the various monomials of the fields needed for defining renormalization for
the integration measure.
This is why we concentrated on such terms in this paper. Eventually, in calculating physical processes, the higher terms with
$\chi$, ${\bar \chi}$ will be needed as well.

It is straightforward to take the $R\rightarrow \infty$ for the term
$m^2_0 \, S_{4d} (H, \chi, {\bar \chi})$ to obtain the flat space limit.
With the metric scaled as indicated in (\ref{0a}), $S_{4d}(H, \chi, {\bar \chi})$
has the dimension of (mass)$^{-2}$, so that $m_0^2$ is retained as such.
But, in the limit of $R\rightarrow \infty$,
the coordinates of the ${\mathbb{R}^4}$ would still be organized into two complex
coordinates. Viewing this as one choice of local complex structure for ${\mathbb{R}^4}$, 
it may be possible to use twistor space and obtain a more symmetric form
as $R \rightarrow \infty$; this is one of the issues for future work.
However, we do emphasize that for any finite value of $R$, no matter how large, 
the term $S_{4d}(H, \chi, {\bar \chi})$ is obtained, and hence it will remain relevant to the question of the mass gap. For this question, it is sufficient to consider the case
$R \gg m^{-1}_R$, but finite.

The need for a dimensional parameter to define nonabelian gauge theories in four dimensions
is certainly not a surprise. We may in fact view this as a nonperturbative version of the standard dimensional transmutation.
It should further be possible to carryout perturbation theory starting from (\ref{52}) -
we will not need gauge-fixing and ghosts - and relate $m^2_R$ to the $\Lambda$-parameter
of QCD.

Going back to the role of the volume element, the two-dimensional version of
$d\mu (\C)$, set into a Hamiltonian formulation, has also been very useful for understanding many features of Yang-Mills theory in three dimensions. The WZW action $S_{wzw}(H)$ from the measure is again crucial for the mass gap in the theory, although it is not directly a 3d-covariant mass term.
In fact, generalizing to the extended supersymmetric cases,
one can show a complete concordance between such terms (or lack thereof) in the functional measure and the results regarding mass gap expected from other independent considerations.
It would be interesting to generalize these considerations to supersymmetric theories in four dimensions by analyzing the measure along the lines of this paper. (There is a small caveat though: Since ${\mathbb{CP}}^2$ is not a spin-manifold, a spin-${\mathbb C}$ structure will have to be used.) We will postpone such an analysis to a future work.

The importance of a mass-like term for Yang-Mills theory in four dimensions
was first emphasized by Cornwall \cite{cornwall} and there have been many attempts to elucidate its origin and implications \cite{CPB}. Our calculation shows a clear and specific realization
of this suggestion.

 The mass-like term in the functional measure, whether for the wave functions at equal time
(as is relevant for a Hamiltonian formulation) or for the Euclidean
spacetime functional integral, provides a cut-off on fluctuations of the low momentum modes
of the fields and this is the key to the mass gap. It is worth emphasizing that this is a general
property of the geometry of the gauge-orbit space and not reliant on any special 
configurations or matter content.

Some of the properties of $S_{4d}(H)$, considered as an action in its own right, are also worthy of a few remarks. The equations of motion for this action give antiself-dual instantons, which are also obviously related to holomorphic vector bundles.
It was in this context that Donaldson originally considered this action. The action $S_{4d}$ was
obtained in \cite{NS} as an attempt to generalize the WZW theory to four dimensions and relate it to the K\"ahler-Chern-Simons theory as a replay of the CS-CFT correspondence in two-three dimensions \cite{witten}. 
As shown in \cite{NS}, and elaborated in \cite{{Los}, {ueno}},
the 4-d theory $S_{4d}$ admits a holomorphically factorized current algebra very similar to the 2-d case. Such theories have also been found in higher dimensional quantum Hall systems
\cite{KNqhe},
and is also realized as the target space dynamics of (world-sheet) $N=2$ heterotic superstrings
\cite{OV}.
Finally, as a small addendum to the remark on the instanton connection, the action
$S_{4d}(H)$ evaluated on instantons is a function of the instanton moduli and 
it would be interesting to see how the integration over the moduli is controlled by the measure
in (\ref{52}).

\bigskip

This work was initiated during a visit to the Niels Bohr International Academy
in April-May, 2012. I thank Poul Damgaard and the members of the academy
for hospitality and D. O'Connell and R. Monteiro for discussions.
I also thank Abhishek Agarwal for useful comments.

This research was also supported by the U.S.\ National Science
Foundation grant PHY-1213380
and by a PSC-CUNY award. 

\end{document}